\documentclass[prb,preprintnumbers,twocolumn,showpacs,nobalancelastpage,noshowpacs]{revtex4}

\usepackage{amsmath,amssymb,amsfonts}
\usepackage{bm,graphicx,color}

\newcommand{\convz}{\ast}

\newcommand{\ret}{{\text{R}}}
\newcommand{\adv}{{\text{A}}}

\newcommand{\les}{<}

\newcommand{\flavorf}{\uparrow}
\newcommand{\flavorc}{\downarrow}

\newcommand{\para}{\text{pm}}
\newcommand{\dia}{\text{dia}}
\newcommand{\reg}{\text{reg}}

\newcommand{\cG}{\mathcal{G}}

\newcommand{\CC}{\mathcal{C}}

\newcommand{\TC}{\text{T}_{\CC}}
\newcommand{\DC}[1]{\partial^\CC_{#1}}

\newcommand{\delC}{\delta_\CC}

\newcommand{\tb}{\bar t}
\newcommand{\tmin}{t_{\text{min}}}
\newcommand{\tmax}{t_{\text{max}}}

\newcommand{\TR}{\text{Tr}}

\newcommand{\vect}[1]{{\bm #1}}
\newcommand{\Vk}{{\vect{k}}}
\newcommand{\Vr}{{\vect{r}}}
\newcommand{\VA}{{\vect{A}}}
\newcommand{\VE}{{\vect{E}}}
\newcommand{\VR}{{\vect{R}}}
\newcommand{\Vq}{{\vect{q}}}
\newcommand{\Vv}{{\vect{v}}}

\newcommand{\Vz}{{\vect{z}}}
\newcommand{\Vj}{{\vect{j}}}
\newcommand{\dA}{{\delta\!\VA}}

\newcommand{\einc}{{E_{\text{0}}}}
\newcommand{\eref}{{E_{\text{refl}}}}
\newcommand{\Ks}{{\Vk\sigma}}

\newcommand{\bmr}{{\bm r}}

\newcommand{\bmA}{{\bm A}}

\newcommand{\bmR}{{\bm R}}

\newcommand{\ket}[1]{|{#1}\rangle}

\newcommand{\expval}[1]{\langle{#1}\rangle}

\newcommand{\myast}{{{}\hspace*{-0.1em}\ast\hspace*{-0.1em}{}}}
\newcommand{\mydagger}{{\dagger}}
\newcommand{\phdagger}{{\phantom{\mydagger}\!}}

\begin{document}

  \title{Theory of time-resolved optical spectroscopy on correlated electron systems}

  \author{Martin Eckstein}

  \author{Marcus Kollar}

  \affiliation{Theoretical Physics III, Center for Electronic
    Correlations and Magnetism, Institute for Physics, University of
    Augsburg, 86135 Augsburg, Germany}

  \date{August 7, 2008}
  
\begin{abstract}
     The real-time dynamics of interacting electrons out of equilibrium
     contains detailed microscopic information about electronically correlated
     materials, which can be read out with time-resolved optical
     spectroscopy.  The reflectivity that is typically measured in
     pump-probe experiments is related to the nonequilibrium optical
     conductivity. We show how to express this quantity in terms of
     real-time Green functions using dynamical mean-field theory.  As
     an application we study the electrical response of the
     Falicov-Kimball model during the ultrafast buildup of the gapped
     phase at large interaction.
   \end{abstract}

  \pacs{PACS}

  \maketitle

  \section{Introduction}
  \label{sec:intro}

  Electronic correlations are known to give rise to highly unusual
  phenomena such as heavy fermion behavior or the Mott metal-insulator
  transitions.\cite{Imada98} In
  recent years a new perspective for this field has been provided by
  various pump-probe spectroscopies, which can directly track the time
  evolution of strongly interacting systems far from equilibrium. For
  example, the dynamics of electrons in the vicinity of a Mott
  metal-insulator transition was investigated using time-resolved
  photoemission spectroscopy\cite{Perfetti06} and time-resolved
  optical
  spectroscopy.\cite{Ogasawara00,Iwai03,Chollet05,Okamoto07,Kuebler07}
  In these experiments, the sample is first excited by an intense
  laser pulse (pump); a second pulse (probe), which comes at a
  controlled time-delay, is then used to characterize the transient
  state by means of photoemission or optical spectroscopy.  Pump-probe
  experiments with femtosecond time-resolution are now commonly
  used for the investigation of dynamics in molecules,\cite{Zewail00}
  metals,\cite{Petek97} and semiconductors.\cite{Axt04} Recent
  development of shorter and shorter pulses has pushed the limiting
  time-resolution below $10$~fs for optical
  frequencies,\cite{Steinmeyer99} and into the attosecond regime for
  pulses in the extreme ultraviolet.\cite{Hentschel01}
  
  For solids it is often a subtle task to distinguish the contribution
  of various degrees of freedom to a specific phenomenon. The Mott
  transition is induced by the Coulomb repulsion between electrons,
  but can occur simultaneously with a change of the lattice structure,
  obscuring the primary origin of the phase transition. In
  time-resolved experiments, however, different degrees of freedom can
  be identified if they evolve on different time
  scales.\cite{Perfetti06,Kuebler07} In particular, the lattice
  usually reacts much slower than the electronic system. Many
  phenomena that are already visible at low time resolution can be
  explained in terms of a two-temperature
  model,\cite{Allen87,Perfetti06} which assumes that the electronic
  system is in thermal equilibrium at any given time, but may have a
  different temperature than the lattice.
  
  On the other hand, pump-probe experiments allow for an investigation
  of the electronic real-time dynamics. For example, two-photon
  photoemission spectroscopy can monitor the ultrafast thermalization
  of a pumped electron gas in metals within several
  $100$~fs.\cite{Fann92a,Petek97} In semiconducting GaAs, the buildup
  of a screened Coulomb interaction in the electron-hole plasma
  created by the photoexcitation of electrons into the conduction band
  has been tracked using time-domain THz spectroscopy.\cite{Huber01}
  In particular the latter experiment probes the true quantum dynamics
  of the state, which can no longer be described by a simple rate
  equation but requires the full many-particle
  Hamiltonian.\cite{Banyai98,Kwong00} It would be very interesting to
  measure the electronic dynamics in strongly interacting systems,
  which may dominate, e.g., the ultrafast buildup of intermediate
  metallic states across insulator-to-metal
  transitions,\cite{Ogasawara00,Iwai03,Okamoto07,Kuebler07} or the
  melting of correlation-induced long-range order after an external
  perturbation.\cite{Chollet05} The goal of this paper is to set up
  the framework for a microscopic description of time-resolved optical
  measurements in such strongly correlated electron systems.
  For time-resolved photoemission spectroscopy, the microscopic 
  description was recently derived in Ref.~\onlinecite{Freericks08b}.

  The microscopic formalism of isolated quantum
  many-body systems out of equilibrium was given independently by Baym
  and Kadanoff,\cite{BaymKadanoff} and Keldysh\cite{Keldysh64} in
  terms of real-time Green functions. It provides the starting point
  for a nonequilibrium perturbation theory,\cite{Rammer86,Haug96} which is
  however bound to fail for strong interactions.  Dynamical mean-field
  theory (DMFT),\cite{Georges96} which becomes exact in the limit of
  infinite spatial dimension,\cite{Metzner89} also applies to the
  non-perturbative regime. DMFT self-consistently maps a lattice model
  onto an auxiliary single-site problem. The equilibrium theory has
  been instrumental in understanding many correlation-induced
  phenomena, such as the Mott transition, both for simple model
  systems,\cite{Georges96} and for real
  materials.\cite{Held06,Kotliar04} Recently, DMFT for nonequilibrium
  has been formulated in the framework of Keldysh
  theory.\cite{Freericks06} It has been used to
  investigate the Falicov-Kimball model\cite{Falicov69,Freericks03}
  under the influence of strong electrical
  fields,\cite{Freericks06,Tsuji08} as well as its relaxation over the
  metal-insulator transition after a sudden change of the interaction
  parameter.\cite{Eckstein08} Similar investigations for the Hubbard
  model still require new techniques for the solution of the effective
  single-site problem. However, promising candidates for this task
  have been developed during the last years.\cite{Anders05,Werner06}

  The main purpose of this paper is to discuss the probe process in
  optical spectroscopy in terms of linear response of a
  nonequilibrium state to an electromagnetic field. For this state,
  which might originate from the application of a pump pulse, the time
  evolution is assumed to be known from DMFT. The response is given by
  the two-time optical conductivity $\sigma(t,t')$, that relates the
  current at time $t$ to electrical fields in the sample at earlier
  times $t'$.\cite{Kindt99} For systems in equilibrium DMFT has
  already been successfully used to understand optical spectroscopy in
  correlated materials.\cite{Rozenberg95} The standard expression for
  the frequency-dependent conductivity $\sigma(\omega)$ in
  DMFT\cite{Pruschke93} is quite simple and contains only
  single-particle Green functions, because vertex corrections to the
  current-current correlation function vanish for isotropic
  systems.\cite{Khurana90,Pruschke93} In this paper we derive an
  expression for the two-time conductivity $\sigma(t,t')$ from
  non-equilibrium DMFT, which turns out to be a direct generalization
  of the equilibrium expression\cite{Pruschke93} to Keldysh language. In
  particular, our derivation shows when the inclusion of vertex
  corrections becomes mandatory in non-equilibrium situations, and
  under which conditions similar simplification occur for
  $\sigma(t,t')$ as for $\sigma(\omega)$.

  We then apply the theory to a simple lattice model for interacting
  electrons in a single band,
  \begin{equation}
  \label{FKM}
  H = \sum_{ij\sigma} V_{ij}^\sigma c_{i\sigma}^\dagger c_{j\sigma}
  + U \sum_{i} n_{i\flavorf} n_{i\flavorc}
  -\sum_{i\sigma} \mu_{\sigma} n_{i\sigma}.
  \end{equation}
  Here $c_{i\sigma}^{(\dagger)}$ are annihilation (creation) operators
  for two species of fermions ($\sigma$ $=$ $\flavorc$,$\flavorf$) on
  lattice site $i$, which interact via a local Coulomb repulsion $U$.
  The first term in (\ref{FKM}) is a tight-binding description of the
  electronic band. Eq.~(\ref{FKM}) is the Hamiltonian of the defines
  the Hubbard model if the hopping $V_{ij}^\sigma$ does not depend on
  the flavor $\sigma$, or the Falicov-Kimball model\cite{Falicov69} if
  one particle species is immobile ($V_{ij}^\flavorf$ $=$ $0$). Both
  models have a rich phase diagram as a function of interaction and
  filling, including metallic, insulation and ordered phases. In the
  presence of electromagnetic fields [with scalar and vector
  potential $\Phi(\Vr,t)$ and $\VA(\Vr,t)$], the hopping amplitudes
  acquire Peierls phase factors \cite{Peierls33,Luttinger51}
  \begin{equation}
  \label{peierls-t}
  V_{ij} = \tilde V_{ij}
  \exp\!\left(
  \frac{ie}{\hbar c}
  \int\limits_{\VR_i}^{\VR_j} \!d\Vr\, \VA(\Vr,t)\right),
  \end{equation}
  and a potential term $-e\sum_{i\sigma} \Phi(\VR_i,t)
  c_{i\sigma}^\dagger c_{i\sigma}$ is added to the Hamiltonian, where
  $-e$ is the charge of an electron. Here and throughout a tilde
  indicates that the quantity is taken in zero external field.

  Nonequilibrium DMFT can potentially model the full pump-probe
  process by including the pump field explicitly in
  Eq.~(\ref{peierls-t}).  In the application of the general result to
  the Falicov-Kimball model we use an idealized nonequilibrium
  situation instead, where the ``pumping'' is an instantaneous event; we
  only have to know the excited state after the pumping, which is taken
  as initial state for the subsequent time evolution. This permits an
  investigation of the relaxation between the various phases.  For
  instance, we can start from a metallic state and follow the
  relaxation in the insulating parameter regime of the Hamiltonian.
  Below we model this situation by a sudden increase of the
  interaction parameter $U$.  We therefore allow for arbitrary time
  dependence of all parameters $U$, $\mu$ and $\tilde{V}^\sigma_{ij}$
  in the Hamiltonian~(\ref{FKM}).

  This paper is outlined as follows. In Section \ref{sec:pump-probe},
  we define the optical conductivity in nonequilibrium experiments,
  and discuss its relation to the reflectivity in time-resolved
  measurements. In Section \ref{sec:DMFT} we shortly review DMFT for
  nonequilibrium. We then derive the nonequilibrium optical
  conductivity in DMFT in Section \ref{sec:opt-dmft}. Finally, in
  Section \ref{sec:optical-results} we apply the theory to the
  Falicov-Kimball model and investigate the response of the system
  during the ultrafast buildup of the gapped phase at large
  interaction.

  \section{Time-resolved optical spectroscopy}
  \label{sec:pump-probe}

  To understand the results of time-resolved optical spectroscopy it
  is necessary to know how weak electromagnetic pulses of finite
  length propagate through the sample, which is not in equilibrium due
  to the applied pump pulse.\cite{Kindt99,Beard01,Schins07} The
  current $\delta\Vj$ is the linear response induced by the probe
  field,
  \begin{align}
  \label{sigma}
  \delta j_\alpha(\Vr,t) =
  \int\limits_{-\infty}^t dt'\, \sigma_{\alpha\beta}(t,t')
  \,\delta E_\beta(\Vr,t'),
  \end{align}
  which defines the optical conductivity $\sigma_{\alpha\beta}(t,t')$
  for samples that are not in equilibrium.  (Here and throughout
  $\alpha$ and $\beta$ are cartesian components of the vectors, and
  repeated indices are summed over.) Note that only the
  response~(\ref{sigma}) is linear in the probe field $\delta
  E_\beta(\Vr,t')$, while arbitrarily strong electric pump fields
  might be acting on the sample.  The wavelength in optical
  spectroscopy is typically much larger than the lattice spacing of
  the sample, so that the linear response relation~(\ref{sigma}) is
  essentially local in space. On the other hand it is not local in
  time, and unless there is a clear separation between the time scales
  that govern the electromagnetic response and the relaxation of the
  nonequilibrium state, $\sigma(t,t')$ depends not only on the
  difference of its time arguments but on both $t$ and $t'$
  separately. Of course $\sigma(t,t')$ is always causal, i.e., it
  vanishes for $t$ $<$ $t'$.

\begin{figure}
  \centerline{\includegraphics[width=6.5cm]{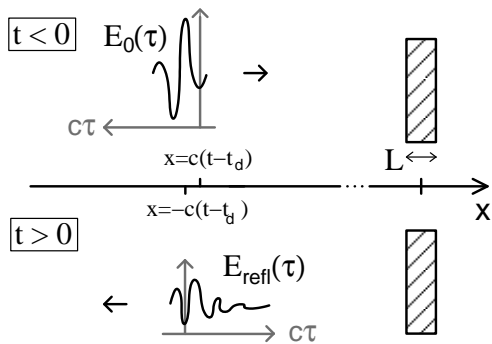}}
  \caption{Time-resolved reflection experiment. For $t$
  $\to$ $-\infty$ a probe pulse $\VE_0(t,x)$ $=$
  $\hat\Vz\einc(t-t_d-x/c)$ propagates in $+x$ direction
  without dispersion (upper panel). The sample is hit at times near
  $t$ $=$ $t_d$, and emits the reflected pulse
  $\VE_{\text{refl}}(t,x)$ $=$ $\hat\Vz\eref(t-t_d+x/c)$, 
  which propagates in $-x$ direction after leaving the sample 
  (lower panel).}
  \label{fig:pump-probe}
  \end{figure}

  Knowledge of $\sigma(t,t')$ is sufficient to calculate the reflected and
  transmitted pulses from Maxwell's equations, assuming that the
  induced current inside the sample is given by
  Eq.~(\ref{sigma}).\cite{Kindt99,Beard01,Schins07} However, the
  relation to measurable quantities is more complicated than for
  samples that are in equilibrium.  To illustrate we this consider a
  typical time-resolved reflection experiment, performed at normal 
  incidence, on a sample that is infinite in the $y$-$z$ plane
  (cf.~Fig.~\ref{fig:pump-probe}).
  Outside the sample light propagates without dispersion, so that we
  may write $\VE_0(t,x)$ $=$ $\hat\Vz\einc(t-t_d-x/c)$ and
  $\VE_{\text{refl}}(t,x)$ $=$ $\hat\Vz\eref(t-t_d+x/c)$ for incident
  and reflected pulses at $x$ $\to$ $-\infty$,
  respectively.  The functions $\einc(\tau)$ and $\eref(\tau)$ are
  centered around $\tau$ $=$ $0$, and $t_d$ is the probe delay. For
  simplicity we assumed cubic symmetry, such that the polarization
  direction $\hat \Vz$ for both pulses is the same.  We then define a
  generalized reflection coefficient $r(t,t')$,\cite{Kindt99}
  \begin{align}
  \label{eref}
  \eref(\tau) = \int\limits_0^\infty ds\,
  \,r(t_d+\tau,t_d+\tau-s)\einc(\tau-s),
  \end{align}
  providing a linear relation between the two pulses.  The full
  two-dimensional time-dependence of $r(t,t')$ can be deduced from
  experiment by suitably choosing the pulse, and measuring at all
  possible pump-probe delays $t_d$. However, if the optical
  conductivity $\sigma(t,t')$ depends on $t$ and $t'$ separately, then
  there is no simple relation to the reflection coefficient
  $r(t,t')$.\cite{Schins07} This is evident from the
  definition~(\ref{eref}), which shows that a sample which is not in
  equilibrium can modulate the pulse frequency.

  From now on we use an approximate form for $r(t,t')$, which is
  valid for reflection from a very thin slab (with thickness $L$ $\to$
  $0$), such that the phase lag between the borders is negligible. In
  this case Maxwell's equations are easily solved, yielding\cite{Kindt99}
  \begin{align}
  \label{reflection}
  r(t,t') = \frac{L}{c}\sigma(t,t').
  \end{align}
  A more realistic description, which takes the finite thickness of
  the sample and its inhomogeneous excited state into account,
  requires the numerical simulation of the pulse
  propagation\cite{Beard01} and of the inverse problem\cite{Schins07} 
  of obtaining $\sigma(t,t')$ from $r(t,t')$.  However, the treatment 
  of such effects is beyond the scope of this paper, the goal of which 
  is to calculate the optical conductivity $\sigma(t,t')$ microscopically
  for an interacting many-body system that is not in equilibrium.

  \section{DMFT for nonequilibrium}
  \label{sec:DMFT}

  DMFT for nonequilibrium usually starts from thermal equilibrium at
  some early time $t$ $=$ $\tmin$.\cite{Freericks06,Eckstein08} For
  $t$ $\geq$ $\tmin$ the system evolves according to the Hamiltonian
  (\ref{FKM}), driven out of equilibrium if the Hamiltonian changes
  with time.  Thermodynamic variables and optical response functions
  are obtained from the retarded, advanced and lesser real-time Green
  functions,
  \begin{subequations}%
    \label{realtimeg}%
    \begin{align}%
      G^\ret_{ij\sigma}(t,t') &=
      -i\Theta(t-t') \expval{\{
        c^{\phantom{\dagger}}_{i\sigma}(t),c^\dagger_{j\sigma}(t')\}}
      \label{gret}\\
      G^\adv_{ij\sigma}(t,t') &=
      i\Theta(t'-t)
      \expval{\{c^{\phantom{\dagger}}_{i\sigma}(t),c^\dagger_{j\sigma}(t')\}}
      \label{gadv}\\
      G^\les_{ij\sigma}(t,t') &=
      i\expval{c^\dagger_{j\sigma}(t')c^{\phantom{\dagger}}_{i\sigma}(t)}.
      \label{glesser}
    \end{align}%
  \end{subequations}%
  (Although retarded and advanced Green functions are in fact related
  by symmetry, both are given here for later reference.) The average
  $\expval{\cdot}$ $=$ $\TR[\rho_0\cdot]$ in Eq.~(\ref{realtimeg}) is
  over initial states at $t$ $=$ $\tmin$, distributed according to the
  grand-canonical density matrix $\rho_0$ $\propto$ $\exp[-\beta
  H(\tmin)]$ at inverse temperature $\beta$. The operators
  $c_{i\sigma}(t)$ $=$ $U(t,\tmin) c_{i\sigma} U(\tmin,t)$ are in
  Heisenberg representation with respect to the full time evolution
  $U(t,t')$ $=$ $T_{\tb} \exp[-i\int_t^{t'}\!\! d\tb\, H(\tb)]$.  Using
  the Keldysh formalism\cite{Keldysh64,Rammer86,Haug96} the Green
  functions (\ref{realtimeg}) are then calculated in terms of a more
  general contour-ordered Green function $G_{ij,\sigma}(t,t')$ $=$
  $-i\expval{\TC c_{i\sigma}^\phdagger(t)c_{j\sigma}^\mydagger(t')}$
  with time arguments on the contour $\CC$ that runs from $\tmin$ to
  some larger time $\tmax$ on the real axis, then from $\tmax$ to
  $\tmin$, and finally to $\tmin-i\beta$ on the imaginary time axis.
  For the retarded, advanced and lesser components one has\cite{Rammer86}
  \begin{subequations}%
    \label{contourrelation}%
    \begin{align}%
      G_{ij\sigma}^\ret(t,t') &= \Theta(t-t')[G_{ij\sigma}^{-+}(t,t')-G_{ij\sigma}^{+-}(t,t')]\\
                              &= G_{ij\sigma}^{++}(t,t')-G_{ij\sigma}^{+-}(t,t')\\
      G_{ij\sigma}^\adv(t,t') &= \Theta(t'-t)[G_{ij\sigma}^{+-}(t,t')-G_{ij\sigma}^{-+}(t,t')]\\
                              &= G_{ij\sigma}^{+-}(t,t')-G_{ij\sigma}^{--}(t,t')\\
      G_{ij\sigma}^\les(t,t') &= G_{ij\sigma}^{+-}(t,t'),
    \end{align}%
  \end{subequations}%
  where the superscripts $\pm$ indicate whether the first and second
  time arguments are on the upper or lower real-time branch of the
  contour, respectively.

  {}From now on we only consider translationally invariant
  nonequilibrium states, i.e., we assume that the Green function
  $G_{ij,\sigma}(t,t')$ depends only on the difference $\VR_i$ $-$
  $\VR_j$, with diagonal Fourier transform $G_\Ks(t,t')$. This assumes
  that the electromagnetic fields do not depend on explicitly on
  position either, which is justified for experiments at optical
  frequencies, as discussed above in Sec.~\ref{sec:pump-probe}.  We
  use a gauge with zero scalar potential $\Phi$, for which electrical
  field is given by $\VE(t)$ $=$ $-\partial_t \VA(t)/c$. The hopping
  amplitude $V_{ij}$ [Eq.~(\ref{peierls-t})] then also depends only on
  the distance $\VR_i-\VR_j$; its Fourier transform $\epsilon_\Ks(t)$
  is given by\cite{Peierls33,Luttinger51}
  \begin{subequations}%
  \begin{align}%
    \epsilon_\Ks(t)
    &=
    \sum_j V_{ij}^\sigma \exp[i\Vk(\VR_j-\VR_i)]
    =
    \tilde \epsilon_{\Vk+\frac{e}{\hbar c}\VA(t),\sigma}
    \,,
    \\
    \label{ek-free}
    \tilde \epsilon_\Ks
    &=
    \sum_j \tilde V_{ij}^\sigma \exp[i\Vk(\VR_j-\VR_i)]
    \,,%
  \end{align}%
  \end{subequations}%
  i.e., $\epsilon_\Ks(t)$ is obtained from the zero-field dispersion
  $\tilde \epsilon_\Ks$ by a time-dependent shift in momentum.

  The interacting contour Green function satisfies the Dyson
  equation\cite{Rammer86,Haug96}
  \begin{align}
  \label{dysonk}
  [(\cG^{-1}_\Ks - \Sigma_\Ks) \convz
  G_\Ks](t,t') = \delC(t,t'),
  \end{align}
  where $\Sigma_\Ks(t,t')$ is the contour self-energy and
  $\cG_\Ks(t,t')$ is the noninteracting Green function,
  whose inverse
  \begin{align}
    \label{g0inverse}
    \cG^{-1}_\Ks(t,t') = \delC(t,t')[i\DC{t} + (\mu_\sigma
    - \epsilon_\Ks(t))/\hbar]
  \end{align}
  can be written as a differential operator on the contour.  Here $(f
  \myast g)(t,t')$ $=$ $\int_\CC d\bar{t}f(t,\bar{t})g(\bar{t},t')$ is
  the convolution of two functions along the contour, $\delC\!(t,t')$
  is the contour delta function [defined by $\int_\CC
  \!d\bar{t}\,f(\bar{t})\delC(\bar{t},t)$ $=$ $f(t)$], and
  $\partial_t^\CC$ denotes the contour derivative.\cite{Freericks06}
  The unique solution of Eq.~(\ref{dysonk}) is determined by
  antiperiodic boundary conditions for the contour Green functions in
  both time arguments.\cite{Rammer86,Haug96}

  The DMFT self-energy is local in space, i.e., $\Sigma_\Ks$ is
  independent of $\Vk$ for a translationally invariant system. This
  approximation becomes exact in the limit of infinite spatial
  dimensions,\cite{Metzner89} both for the equilibrium self-energy and
  the Keldysh self-energy.\cite{Freericks06} In DMFT the local
  self-energy $\Sigma_\sigma(t,t')$ and the local Green function
  $G_{\sigma}(t,t')$,
  \begin{align}
    \label{gloc}
    G_{\sigma}(t,t') \equiv
    G_{ii\sigma}(t,t') =
    \frac{1}{N}
    \sum_\Vk G_\Ks(t,t')
   \,,
  \end{align}
  ($N$ is the number of lattice sites in the sample) 
  are determined from an auxiliary problem in which the degrees of freedom
  at a single lattice site $i$ are coupled to some unknown
  environment. The latter must be determined self-consistently, by
  solving the auxiliary problem together with the Dyson
  equation~(\ref{dysonk}).
  As the precise form of the local problem in terms of its many-body action
  does not enter into the derivation of the electromagnetic response below,
  we refer to previous work for further details.\cite{Freericks06,Eckstein08} 


  \section{Optical conductivity in DMFT}
  \label{sec:opt-dmft}

  The current operator for the Hamiltonian (\ref{FKM}) is
  defined\cite{footnote1,Shastry90,Scalapino92} by the relation
  $\Vj(\Vr)$ $=$ $-c\delta H / \dA(\Vr)$.  Using Eq.~(\ref{peierls-t}),
  we obtain the current in the long wave-length limit as
  \begin{subequations}%
    \begin{align}
      \label{j-expval}
      \expval{\Vj(t)}
      &=
      \left\langle
        \frac{1}{V}
        \int d^d\Vr \Vj(\Vr)e^{i\Vq\Vr}
      \right\rangle_{\!\Vq\to0}
      \,,
      \\
      &=
      \frac{ie}{V}
      \sum_\Ks\Vv_\Ks(t)
      G_\Ks^<(t,t)
      \,,\label{j-gf}
    \end{align}%
    the current vertex is given by 
    \begin{align}%
      \label{jvertex}
      \Vv_\Ks(t)
      =
      \hbar^{-1}\partial_\Vk \epsilon_\Ks(t)
      =
      \hbar^{-1}\partial _\Vk \tilde \epsilon_{\Vk+\frac{e}{\hbar c}\VA(t),\sigma}
      \,.
    \end{align}%
  \end{subequations}%
  and $V$ is the volume of the sample.  Although the response to
  arbitrarily strong fields is described by DMFT\cite{Freericks06},
  here we are interested in the linear current response to a weak
  probe field.  We define the susceptibility
  \begin{align}
    \label{chi}
    \chi_{\alpha\beta}(t,t')
    =
    \delta \expval{j_\alpha(t)}/A_\beta(t').
  \end{align}
  In the chosen gauge with $\VE(t)$ $=$ $-\partial_t \VA(t)/c$, the
  susceptibility $\chi_{\alpha\beta}(t,t')$ is related to the optical
  conductivity $\sigma_{\alpha\beta}(t,t')$ [Eq.~(\ref{sigma})] by
  \begin{align}
    \label{chi2sigma}
    \sigma_{\alpha\beta}(t,t')
    &=
    -c\int\limits_{t'}^\infty d\tb \,\chi_{\alpha\beta}(t,\tb).
  \end{align}
  The susceptibility (\ref{chi}) is related to the current-current
  correlation function, which can be evaluated in analogy to the
  equilibrium case.\cite{Pruschke93} Here we prefer to take the
  derivative of~(\ref{j-gf}) directly, where the vector potential
  enters both in the vertex $\Vv_\Ks(t)$ and in the Green function
  $G_\Ks^<(t,t)$. This yields the diamagnetic and paramagnetic
  contributions to the susceptibility,
  \begin{subequations}%
    \begin{align}%
      \chi_{\alpha\beta}(t,t')
      &=
      \chi^{\dia}_{\alpha\beta}(t,t')
      +
      \chi^{\para}_{\alpha\beta}(t,t')
      \,,
      \\
      \chi^{\dia}_{\alpha\beta}(t,t')
      &=
      \frac{ie}{V}
      \sum_\Ks\frac{\delta v^\alpha_\Ks(t)}{\delta A_\beta(t')}\,
      G_\Ks^<(t,t)
      \,,\label{chi-dia1}
      \\
      \chi^{\para}_{\alpha\beta}(t,t')
      &=
      \frac{ie}{V}
      \sum_\Ks v^\alpha_\Ks(t)\,
      \frac{\delta G_\Ks^<(t,t)}{\delta A_\beta(t')}
      \,.\label{chi-para1}
    \end{align}%
  \end{subequations}%
  The paramagnetic contribution can be found from a variation of the
  lattice Dyson equation~(\ref{dysonk}),
  \begin{align}
  \label{deltagk}
  \delta G_\Ks =
  -
  G_\Ks \convz [ \delta\cG_\Ks^{-1} - \delta\Sigma_{\sigma}]
  \convz G_\Ks.
  \end{align}

  Some simplifications occur in the absence of anisotropies. We note
  that the second term in~(\ref{deltagk}), containing the
  $\Vk$-independent self-energy, does not contribute to the $\Vk$-sum
  in Eq.~(\ref{chi-para1}) if, under inversion of $\Vk$, (i) $G_\Ks$
  is symmetric and (ii) the vertex $\Vv_\Ks$ is antisymmetric. These
  conditions are met by an isotropic system without external fields,
  and are therefore generally valid for systems with inversion
  symmetry in equilibrium.\cite{Khurana90} However, the isotropy may
  be lost when an initially isotropic system is driven out of
  equilibrium, e.g., when a current is induced by the electrical pump
  field. Furthermore, the vertex (\ref{jvertex}) is no longer
  antisymmetric when an electrical field is present in addition to the
  probe field, i.e., when the paramagnetic
  susceptibility~(\ref{chi-para1}) is evaluated at $\VA$ $\neq$ $0$.
  Experimentally these anisotropic effects in otherwise isotropic
  systems show up as a dependence of the signal on the relative
  polarization of pump and probe pulses.  However, when the anisotropy
  is caused entirely by the pump pulse, the inversion symmetry of
  $G_\Ks \convz$ $\delta\Sigma_{\sigma}\convz$ $ G_\Ks$ can be
  restored by averaging over the pump pulse polarization. Then this
  term again drops out in (\ref{chi-para1}), provided that
  $\Vv_\Ks(t)$ is antisymmetric (i.e., $\VA(t)$ $=$ $0$).  In order to
  study such anisotropic effects, vertex corrections contained in
  $\delta\Sigma_\sigma$ must be taken into account (even for cubic
  lattices), by solving a Bethe-Salpeter equation on the Keldysh
  contour, with the irreducible vertex function
  $\delta\Sigma_\sigma/\delta G_\sigma$ from the auxiliary single-site
  problem as input.

  In the following we only consider the completely isotropic
  relaxation between homogeneous phases, such that the vertex
  corrections $\delta \Sigma_\sigma$ can be disregarded.
  Eq.~(\ref{chi-para1}) is evaluated at zero field, so that only the
  first term $\delta F_\Ks(t_1,t_2)$ $=$ $-[G_\Ks \convz
  \delta(\cG_\Ks^{-1}) \convz G_\Ks](t_1,t_2)$ contributes to $\delta
  G_\Ks$ in Eq.~(\ref{deltagk}). This corresponds to keeping only the
  elementary bubble diagram for the current-current correlation
  function.\cite{Pruschke93} The two convolutions in $\delta
  F_\Ks(t_1,t_2)$ collapse to a single one because $[\delta
  \cG_\Ks^{-1}](t,t')$ $\propto$ $\delC(t,t')$.  In order to obtain
  $\delta G_\Ks^\les(t,t)$ we take $t_1$ $=$ $t$ and $t_2$ $=$
  $t$ on the upper and lower branch of the contour, respectively
  [cf.~Eq.~(\ref{contourrelation})]. The contour integral is then
  transformed into an integral along the real axis,
  \begin{multline}
  \delta F_\Ks(t_+,t_-) = \frac{e}{\hbar c}
  \int\limits_{-\infty}^{\infty}\!d\tb\, \Vv_\Ks(\tb)\dA(\tb)
  \\
  \times\,
  [
  G^{++}_\Ks(t,\tb) G^{+-}_\Ks(\tb,t)-
  G^{+-}_\Ks(t,\tb) G^{--}_\Ks(\tb,t)
  ]
  \end{multline}
  from which the optical conductivity $\sigma_{\alpha\beta}(t,t')$ can
  be read off.  From Eq.~(\ref{contourrelation}), together with the
  relations $G^\les_\Ks(t,t')$ $=$ $-G^\les_\Ks(t',t)^*$ and
  $G^\ret_\Ks(t,t')$ $=$ $G^\adv_\Ks(t',t)^*$, we finally obtain the
  paramagnetic susceptibility
  \begin{subequations}%
    \label{chifinal}%
    \begin{align}%
      \label{chi-para}%
      \chi^{\para}_{\alpha\beta}(t,t') = 
      \!-2\chi_0\!
      \sum_\Ks
      \tilde v_\Ks^\alpha
      \tilde v_\Ks^\beta
      \text{Im}[G^\ret_\Ks(t,t')G^\les_\Ks(t',t)],
    \end{align}%
    where $\chi_0$ $=$ $e^2/(V\hbar c)$ and $\tilde\Vv_\Ks$ $=$
    $\partial_\Vk\tilde\epsilon_{\Vk,\sigma}/\hbar$.  The diamagnetic
    contribution follows directly from (\ref{jvertex}):
    \begin{align}%
      \label{chi-dia}%
      \chi^{\dia}_{\alpha\beta}(t,t') = 
      \frac{i\chi_0}{\hbar}
      \delta(t-t')\!
      \sum_\Ks
      (\partial_{\Vk_\alpha}\!
      \partial_{\Vk_\beta}
      \tilde\epsilon_\Ks)
      G^\les_\Ks(t,t)
      \,.
    \end{align}%
  \end{subequations}%
  Eqs.~(\ref{chi2sigma}) and (\ref{chifinal}) constitute our final
  DMFT expressions for the optical conductivity (provided that
  anisotropic effects are disregarded, as discussed above).

  The optical conductivity (\ref{chi2sigma}) can be written as
  \begin{align}
    \sigma_{\alpha\beta}(t,t')
    =
    [\sigma_{\alpha\beta}^\reg(t,t')
    +D_{\alpha\beta}(t)]\Theta(t-t')
    \,,
  \end{align}
  i.e., it splits into its regular part
  \begin{align}
    \label{sigmareg}
    \sigma_{\alpha\beta}^\reg(t,t')
    =
    c\int\limits_{-\infty}^{\phantom{'}t'} \!d\tb \,\chi_{\alpha\beta}^\para(t,\tb)
    \,,
  \end{align}
  which vanishes in the limit $t'$ $\to$ $-\infty$, and the Drude
  contribution
  \begin{subequations}%
    \label{drude}%
    \begin{align}%
      D_{\alpha\beta}(t) &\equiv
      \lim_{t'\to -\infty} \sigma_{\alpha\beta}(t,t')
      \\
      &=
      \sigma_{\alpha\beta}^\dia(t)
      -
      c\int\limits_{-\infty}^{t} \!d\tb \,\chi_{\alpha\beta}^\para(t,\tb),
    \end{align}%
  \end{subequations}%
  which does not depend on the time difference at all. In the latter
  expression, $\sigma_{\alpha\beta}^\dia(t)$ $=$
  $-c\int_{-\infty}^{\infty} dt'\,\chi_{\alpha\beta}^\dia(t,t')$ is the
  weight of the delta function in Eq.~(\ref{chi-dia}). A finite Drude
  contribution $D_{\alpha\beta}(t)$ $\neq$ $0$ indicates perfect
  metallic behavior, because it gives rise to a delta function at zero
  frequency in the partially Fourier-transformed optical conductivity
  \begin{subequations}%
    \label{sigma-ft}%
    \begin{align}
      \label{sigma-tw}
      \tilde\sigma_{\alpha\beta}(t,\omega)
      &=
      \int\limits_0^\infty \!ds\, e^{i(\omega+i0) s} \sigma_{\alpha\beta}(t,t-s)
      \\
      \label{sigma-tx2}
      &=
      \tilde\sigma^\reg_{\alpha\beta}(t,\omega)
      +
      \frac{iD_{\alpha\beta}(t)}{\omega+i0}
      \,.
    \end{align}%
  \end{subequations}%
  Note that Eqs.~(\ref{chifinal}) and~(\ref{sigmareg}) can be checked
  by inserting equilibrium Green functions
  \begin{subequations}%
    \begin{align}%
      G^\ret_\Ks(t,t') &= -i\Theta(t-t')\int \!d\omega \,A_\Ks(\omega) e^{i\omega (t'-t)}
      \,,\\
      G^\les_\Ks(t,t') &= i\int \!d\omega \,A_\Ks(\omega)f(\omega) e^{i\omega (t'-t)}
      \,,
    \end{align}%
  \end{subequations}%
  with the spectral function $A_\Ks(\omega)$ $=$
  $-\text{Im}[G^\ret_\Ks(\omega$ $+$ $i0)]/\pi$ and the Fermi function
  $f(\omega)$ $=$ $1/(1+e^{\beta\omega})$, which depend only on time 
  differences, into Eq.~(\ref{sigma-ft}).
  Then the well-known expression for the regular part of the optical
  conductivity in equilibrium,\cite{Pruschke93}
  \begin{multline}
    \label{sigma-eq}
    \text{Re}\, \sigma^\reg_{\alpha\beta}(\omega) =
    \pi c\chi_0
    \sum_\Ks
    \tilde v_\Ks^\alpha
    \tilde v_\Ks^\beta
    \,\times
    \\
    \int\limits_{-\infty}^{\infty} \!d\omega'\,
    \frac{A_\Ks(\omega')A_\Ks(\omega+\omega')
      [f(\omega')-f(\omega+\omega')]}{\omega},
  \end{multline}
  is recovered.

  \section{Pump-probe spectroscopy on the  Falicov-Kimball model}
  \label{sec:optical-results}

  \subsection{The Falicov-Kimball model in nonequilibrium}

  In the remaining part of this paper we focus on a specific
  electronic model, the Falicov-Kimball model. This lattice model
  describes itinerant ($\flavorc$) electrons and immobile ($\flavorf$)
  electrons that interact via a repulsive local interaction
  $U$.\cite{Falicov69} The Hamiltonian is given by Eq.~(\ref{FKM})
  with $V_{ij}^\flavorf$ $=$ $0$. The Falicov-Kimball model has been
  an important benchmark for the development of DMFT in equilibrium,
  because the effective single-site problem for the mobile particles
  is quadratic and can be solved exactly.\cite{Brandt89} It currently
  plays a similar role for nonequilibrium
  DMFT,\cite{Freericks06,Eckstein08,Tsuji08} in particular since no appropriate
  real-time impurity solver is yet available for the Hubbard model.
  In spite of its apparent simplicity the Falicov-Kimball model has a
  rich phase diagram containing metallic, insulating, and
  charge-ordered phases.\cite{Freericks03} In the following we fix the
  filling of both particle species ($n_\flavorc$ $=$ $n_\flavorf$ $=$
  $1/2$), and consider only the homogeneous phase without symmetry
  breaking. This phase undergoes a metal-insulator transition at a
  critical interaction $U$ $=$
  $U_c$,\cite{Brandt89,vanDongen90,vanDongen92} from the gapless phase
  at $U$ $<$ $U_c$ to the gapped phase at $U$ $>$ $U_c$.

  Below we assume that the system is prepared in thermal equilibrium
  for times $t$ $<$ $0$. Then the interaction parameter $U$ is changed
  abruptly at $t$ $=$ $0$.  In this way we study the relaxation of the
  system in the insulating parameter regime, starting from a weakly
  correlated state ($U$ $<$ $U_c$).  This mimics an experiment similar 
  to the one described in Ref.~\onlinecite{Huber01}, where the buildup 
  of a weakly correlated state is studied with time-resolved spectroscopy, 
  starting from an
  uncorrelated state of electrons just after their excitation into an
  empty conduction band.  Note that in this interpretation the state
  of the conduction band {\em immediately after the pump pulse} is the
  initial state for the relaxation process.

  The relaxation dynamics after such an interaction quench was
  recently investigated with DMFT using the exact Green functions
  $G_{\Vk\flavorc}(t,t')$ of the mobile particles.\cite{Eckstein08}
  However, only thermodynamic observables were discussed in
  Ref.~\onlinecite{Eckstein08}, with a special focus on their steady
  state value in the long-time limit. Here we consider instead
  hypothetical time-resolved experiments that are performed on the
  system during relaxation, i.e., we use the Green functions from
  Ref.~\onlinecite{Eckstein08} to evaluate the optical conductivity
  from Eq.~(\ref{chifinal}).  Momentum summations in (\ref{chifinal})
  are performed for a hypercubic lattice, taking the dispersion
  $\tilde\epsilon_\Vk$ to be that of a semielliptic density of
  states,\cite{Bluemer03} $\rho(\epsilon)$ $=$ $(2/\pi W^2)
  \sqrt{W^2-\epsilon^2}$ (cf.~App.~\ref{app:2}).  The half-bandwidth
  $W$ $=$ $2$ sets the energy scale, such that the critical
  interaction is $U_c$ $=$ $W$ $=$ $2$.

  \subsection{Optical conductivity and reflected electrical field}

\begin{figure}
  \centerline{\includegraphics[width=9cm]{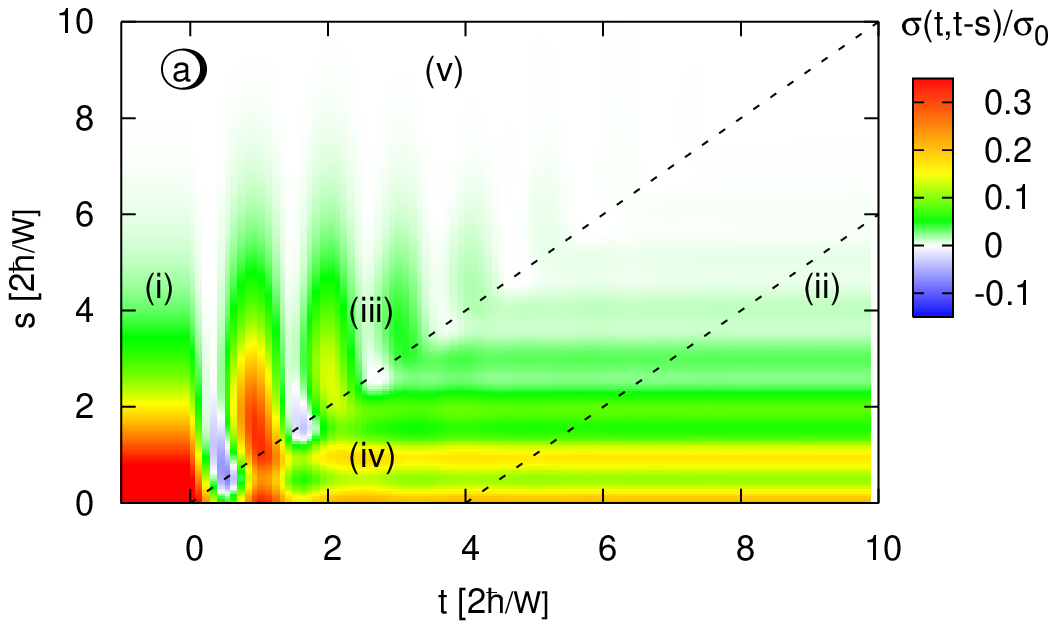}}
  \vspace*{-4mm}
  \centerline{\includegraphics[width=4.2cm]{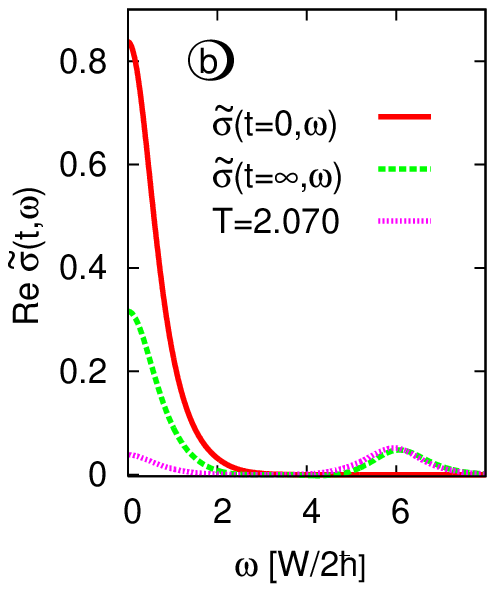}%
  \hspace*{1mm}%
  \includegraphics[width=4.2cm]{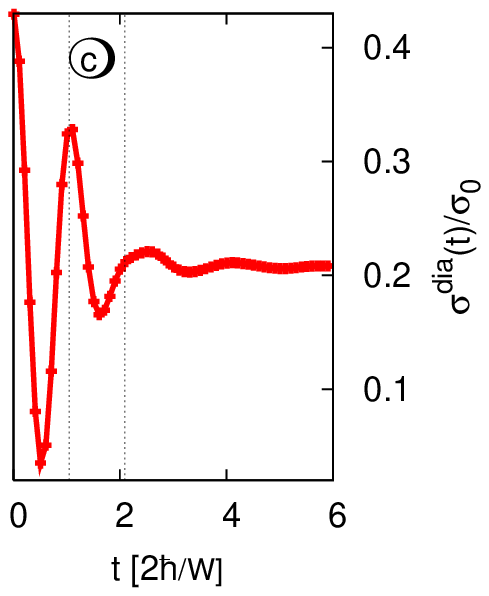}}
  \caption{(a) Optical conductivity $\sigma(t,t-s)$ for the quench from
  the ground state at $U$ $=$ $1$ (initial temperature $T$ $=$ $0$) to
  $U$ $=$ $6$ ($n_f$ $=$ $n_c$ $=$ $1/2$, half-bandwidth $W$ $=$ $2$).
  The unit of the conductivity is $\sigma_0$ $=$ $N a^2 e^2 W /(2\hbar^2 V)$,
  where $a$ is the lattice constant.
  In the region above the upper dashed line, $t-s$ $<$ $0$.  Below the
  lower dashed line the relaxation is essentially complete. (b)
  Fourier transform (\ref{sigma-tw}) of the optical conductivity in
  the initial and final stationary state, and for an equilibrium state
  at $U$ $=$ $6$, with the same excitation energy relative as the
  final state ($T$ $=$ $2.070$). (c) Diamagnetic contribution
  (\ref{chi-dia}) to the susceptibility.}
  \label{fig:sigmaU1-6}
  \end{figure}

  We study relaxation far in the insulating regime ($U$ $=$ $6$),
  starting from an initial metallic state ($U$ $=$ $1$). The optical
  conductivity $\sigma(t,t-s)$ for this case is shown in
  Fig.~\ref{fig:sigmaU1-6}a as a function of $t$ and $s$.
  There are five regions [(i) to (v)] in this plot that we want to discuss
  in detail.

  In regions (i) [$t$ $<$ $0$] and (ii) [$t$ $\to$ $\infty$],
  $\sigma(t,t-s)$ depends only on the time-difference $s$, indicating
  that the system is in a stationary state. For (i) this is the
  initial equilibrium state, and for (ii) it corresponds to the final
  steady state.\cite{Eckstein08} The Fourier transformation
  (\ref{sigma-tw}) of the conductivity exhibits a broad peak at
  $\omega$ $=$ $0$, both for the initial state
  [$\tilde\sigma(t=0,\omega)$] and the final state
  [$\tilde\sigma(t=\infty,\omega)$] (cf.~Fig.~\ref{fig:sigmaU1-6}b).
  This clear indication of metallic behavior of the final state may
  seem surprising, since the interaction is far above the critical
  interaction $U_c$.  However, a finite DC conductivity should be
  expected because the final state is highly excited with respect to
  the ground state at $U$ $=$ $6$. In fact, the excitation energy
  corresponds to an effective temperature $T$ $=$ $2.070$, for
  which the equilibrium DC conductivity $\sigma(0)$ is already quite
  sizable even at $U$ $=$ $6$ (dotted curve in
  Fig.~\ref{fig:sigmaU1-6}b). However, $\sigma(0)$ is still
  considerably lower compared to $\tilde\sigma(t=\infty,0)$.  This is
  a signature of the incomplete relaxation in the Falicov-Kimball
  model: The system does not relax to thermal equilibrium, but reaches
  a non-thermal steady state, as shown in Ref. \onlinecite{Eckstein08}
  for thermodynamic quantities.  In the present context we find that
  the electromagnetic response of the non-thermal final state combines
  some features of the insulating state (a peak around $\omega$ $=$
  $6$ due to excitations across the gap) with a sizable DC
  conductivity. Full thermalization is expected only due to coupling
  to further degrees of freedom or further hopping or interaction
  terms that are not contained in~(\ref{FKM}).

  For $t-s$ $<$ $0$ and $t$ $>$ $0$ [region (iii) in
  Fig.~\ref{fig:sigmaU1-6}a], $\sigma(t,t-s)$ determines the current
  {\em after the pumping at $t$ $=$ $0$} caused by an electrical field
  applied to the sample {\em before the pumping}. It thus measures a
  combination of the electromagnetic response of the initial state and
  the subsequent decay of the induced current for $t$ $>$ $0$.  By
  contrast, in region (iv) in Fig.~\ref{fig:sigmaU1-6}a it describes
  the response of the nonequilibrium state alone, and hence gives
  direct insight into various relaxation processes.  True
  nonequilibrium dynamics can be observed only when both $t-s$ and $t$
  are smaller than some relaxation time $\tau_{\text{stat}}$, after
  which the response is stationary, i.e., when $\sigma(t,t-s)$ depends
  on $s$ only. In the present case the relaxation is virtually
  complete after only a few times of the inverse half-bandwidth
  ($\tau_{\text{stat}}$ $\approx$ $8/W$ $=$ $4$, below the lower
  dotted line in Fig.~\ref{fig:sigmaU1-6}a).  Therefore the relaxation
  time and the time scales of the electromagnetic response, which is
  set by the decline of $\sigma(t,t-s)$ for $s$ $\to$ $\infty$,
  apparently have the same order of magnitude.

  In spite of this very fast relaxation nontrivial transient behavior
  can be observed before the stationary state is reached.  Consider
  $\sigma(t,t-s)$ at $s$ $=$ $0$, which traverses almost two damped
  oscillation cycles with an approximate period $2\pi\hbar/U$ before
  reaching its final value (Fig.~\ref{fig:sigmaU1-6}c). Recall that
  $\sigma(t,t)$ is given by the delta function weight $\sigma^\dia(t)$
  of the diamagnetic susceptibility (\ref{chi-dia})
  [cf.~Eqs.~(\ref{sigmareg}) and (\ref{drude})].  These oscillations are
  the hallmark of dynamics that are dominated by a Hubbard-type
  density interaction such as $U \sum_i n_{i\flavorf} n_{i\flavorc}$.
  In fact, when the Hamiltonian is given only by this interaction
  term, the time evolution-operator $\exp[itU \sum_i n_{i\flavorf}
  n_{i\flavorc}]$ itself is time-periodic,\cite{Greiner02} and
  oscillations should therefore be visible in all nonlocal quantities.
  These so-called collapse-and-revival oscillations were first
  observed and described in experiments with ultra-cold atomic
  gases,\cite{Greiner02} where the Hamiltonian of the system can be
  designed in a controlled way.

  Finally we note that the conductivity $\sigma(t,t-s)$ vanishes in
  the limit $s \to$ $\infty$ , i.e., the Drude weight (\ref{drude})
  vanishes for all times [region (v) in Fig.~\ref{fig:sigmaU1-6}a].
  This is well known for the Falicov-Kimball model in
  equilibrium:\cite{Freericks03} unlike in the Hubbard
  model,\cite{Scalapino92} the mobile particles do not form a perfect
  metal even at $T$ $=$ $0$ because of the disordered background of
  immobile particles.  Mathematically the vanishing of
  $D_{\alpha\beta}(t)$ is due to the cancellation of the two terms in
  (\ref{drude}). Since each of them has a nontrivial time dependence
  (cf.  Fig.~\ref{fig:sigmaU1-6}c), this cancellation represents a
  strong check for our numerical evaluation of the conductivity.

\begin{figure}
  \centerline{\includegraphics[width=9cm,bb=50 50 372 125]{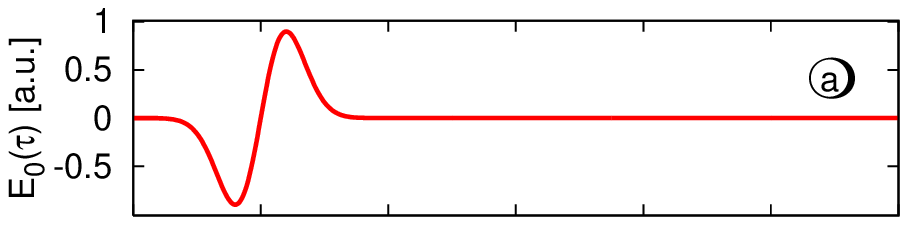}}
  \centerline{\includegraphics[width=9cm,bb=50 55 372 244,clip=true]{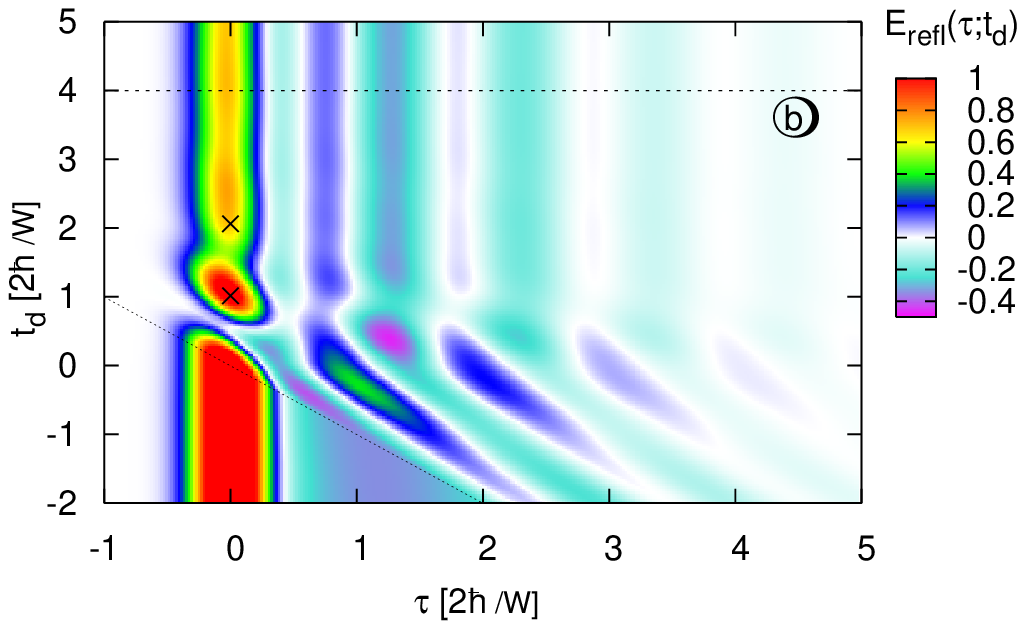}}
  \caption{Result of an idealized spectroscopy experiment
    (cf.~Fig.~\ref{fig:pump-probe}): (a) Incident pulse. (b) Reflected
    pulse $\eref(\tau;t_d)$ [from Eqn. (\ref{eref}) and
    (\ref{reflection})], for a delay $t_d$ of the incident pulse with
    respect to the start of the relaxation at $t$ $=$ $0$. The region
    below the diagonal dotted line is not influenced by the quench at
    all. Above the horizontal dotted line ($t_d$ $\gtrsim$
    $\tau_{\text{stat}}$ $=$ 4) the reflected signal is converged.
    For $t_d$ $<$ $\tau_{\text{stat}}$, at least one revival peak at
    $t_d$ $=$ $2\pi\hbar/U$ is clearly visible (crosses).}
  \label{fig:eref}
  \end{figure}

  To illustrate the relation of the optical conductivity
  to time-resolved THz experiments, we  use the simple
  expression~(\ref{reflection}) for the reflection coefficient,
  and calculate the reflected field $\eref(\tau;t_d)$ according
  to the definition (\ref{eref}), using a single cycle incident
  pulse $\einc(\tau)$ $=$ $\sin(\tau)\exp(-2\tau^2)$. The result is shown
  in Fig.~\ref{fig:eref}.
  For short delay times $t_d$ between the incident pulse and the pump-event 
  at $t$ $=$ $0$, the profile of the reflected field depend strongly on $t_d$.
  On the other hand, for times $t_d$ $\gtrsim$ $\tau_{stat}$, the relaxation is essentially
  complete, and $\eref(\tau)$ has developed a longer oscillating tail.  This general behavior
  is also seen in the experiment of Ref.~\onlinecite{Huber01}. 
  In Fig.~\ref{fig:eref} the oscillations in $\eref(\tau)$
  as a function of $\tau$ are characteristic of the gap in the final state.
  Furthermore, the above-mentioned transient $2\pi\hbar/U$-periodic oscillations 
  are visible in the $t_d$ dependence  of the reflected field $\eref(\tau)$  at 
  small $\tau$.

  \section{Conclusion}
  
  In this paper we generalized the familiar equilibrium expression for
  the optical conductivity in DMFT to the linear electromagnetic
  response of a nonequilibrium state. We find that the two-time
  optical conductivity $\sigma(t,t')$, which is probed in
  time-resolved optical spectroscopies, can be expressed in terms of
  electronic real-time Green functions [see Eqs.~(\ref{chi2sigma}) and
  (\ref{chifinal})], which can be obtained from the DMFT solution. The
  expression for $\sigma(t,t')$ is completely general. Only
  anisotropic effects are disregarded that would lead to a dependence
  of the signal on the relative polarization direction of pump and
  probe pulses, i.e., averaging over the pump-probe direction is
  assumed.
  
  As a first application we have applied the theory to a hypothetical
  pump-probe experiment on the Falicov-Kimball model.  The pumping out
  of equilibrium was modelled by a sudden change in the interaction
  parameter, after which an electrical field pulse probes the
  relaxation between metallic and insulating phases. We observe very
  fast relaxation with a relaxation time comparable to the inverse
  bandwidth, such that no clear separation of the time scales occurs
  between the intrinsic relaxation and electromagnetic response.
  Moreover, the two-time optical conductivity reveals transient
  oscillations in the response on a shorter time scale on the order of
  the inverse interaction. These collapse-and-revival oscillations are
  expected to be very robust, e.g., for different densities.  Using
  time-resolved spectroscopy it may thus be possible to observe this
  phenomenon, which is known from experiments with ultracold atoms in
  optical lattices, in the relaxation of correlated electrons in
  solids as well.
  
  In the future, it should become feasible to solve the DMFT equations
  also for the Hubbard model in nonequilibrium. This will provide
  important insight into the dynamics of the pumped Mott insulator at 
  short time-scales.


  \section*{Acknowledgements}
  We thank Dieter Vollhardt for valuable discussions.
  M.E.\ acknowledges support by Studienstiftung des Deutschen Volkes.
  This work was supported in part by the SFB 484 of the Deutsche
  Forschungsgemeinschaft.


  \begin{appendix}
  
  \section{Momentum summations}
  \label{app:2}
  
  For the homogeneous and isotropic relaxation without external
  fields discussed in Section \ref{sec:optical-results}, the
  evaluation of momentum sums is performed along the same lines as in
  equilibrium:\cite{Pruschke93} Because the DMFT self-energy
  $\Sigma_\Ks$ $\equiv$ $\Sigma_\sigma$ is local, the momentum $\Vk$
  enters the DMFT equations (\ref{dysonk})-(\ref{gloc}) only via the
  single-particle energy $\tilde\epsilon_\Ks\sigma$
  [Eq.~(\ref{ek-free})], i.e., $G_\Ks(t,t')$ $\equiv$
  $G_{\tilde\epsilon_\Ks\sigma}(t,t')$ in zero field.\cite{Eckstein08} The $\Vk$
  sums in Eq.~(\ref{gloc}), (\ref{chi-para}), and~(\ref{chi-dia}) can
  then be reduced to integrals over a single energy
  variable\cite{Turkowski05,Freericks06} by introducing the local density of
  states
  \begin{equation}
  \label{ldos}
  \rho_\sigma(\epsilon) = 
  \sum_\Vk 
  |\expval{i|\Ks}|^2 
  \delta(\epsilon-\tilde\epsilon_\Ks)
  \end{equation}
  and the dispersion function 
  \begin{equation}
  \label{D-function}
  D_{\alpha\beta}^\sigma(\epsilon) 
  =
  \frac{1}{N}\sum_\Vk 
  \delta(\epsilon-\tilde\epsilon_\Ks)
  \tilde v_\Ks^\alpha 
  \tilde v_\Ks^\beta.
  \end{equation} 
  In Eq.~(\ref{ldos}), $\ket{\Ks}$ is the single particle state of the
  hopping matrix $\tilde V_{ij}^\sigma$; for a Bravais lattice one has
  $|\expval{i|\Ks}|^2$ $=$ $1/N$. For any function $g(\epsilon)$ we
  thus obtain the relations
  \begin{equation}
  \frac{1}{N}\sum_\Vk g(\tilde\epsilon_\Ks)
  =
  \int\limits_{-\infty}^{\infty} \!d\epsilon\,\rho_\sigma(\epsilon)\,g(\epsilon)
  \end{equation}
  in Eq.~(\ref{gloc}), and 
  \begin{align}
  \frac{1}{N}
  \sum_\Vk
  \tilde v_\Ks^\alpha
  \tilde v_\Ks^\beta\,
  g(\tilde\epsilon_\Ks)
  &=
  \int\limits_{-\infty}^{\infty} \!d\epsilon
  \,D_{\alpha\beta}^\sigma(\epsilon)\,g(\epsilon)
  \\
  \frac{1}{\hbar^2 N}
  \sum_\Vk
  (\partial_{\Vk_\alpha}\!
  \partial_{\Vk_\beta}
  \tilde\epsilon_\Ks)\,
  g(\tilde\epsilon_\Ks)
  &=
  \int\limits_{-\infty}^{\infty} \!d\epsilon\;
  [\partial_\epsilon D_{\alpha\beta}^\sigma(\epsilon)]\,g(\epsilon)
  \end{align}
  in Eqs.~(\ref{chi-para}) and~(\ref{chi-dia}).  Here the last
  relation is proven using partial integration and the identity
  $\tilde\Vv_\Ks\partial_\epsilon \delta(\epsilon-\tilde\epsilon_\Ks)$
  $=$ $-\partial_\Vk \delta(\epsilon-\tilde\epsilon_\Ks)$.
  
  In this work we use a semielliptic density of states,
  $\rho_\flavorc(\epsilon)$ $=$ $(2/\pi W^2)\sqrt{W^2-\epsilon^2}$ for
  the mobile particles in the Falicov-Kimball model, which leads to a
  simple self-consistency condition for the auxiliary single-site
  problem.\cite{Eckstein08} In the limit of infinite coordination
  number, this density of states is obtained for nearest-neighbor
  hopping on the Bethe lattice, but also for a particular choice of
  longer range hopping amplitudes on the hypercubic
  lattice.\cite{Bluemer03} In the latter case one
  obtains\cite{Bluemer03}
  \begin{multline}
  D_{\alpha\beta}^\flavorc(\epsilon)
  =
  \delta_{\alpha\beta}
  \frac{W a^2}{4\hbar^2\sqrt{1-(\epsilon/W)^2}} \,\times
  \\
  \!\!\!\exp\!\left[
    -2\,
    \text{erf}^{-1}
    \!\left(
      \frac{
        \epsilon\sqrt{1-(\epsilon/W)^2}+W\sin^{-1}(\epsilon/W)
      }
      {\pi W/2}
    \right)^{\!\!2\,}
  \right]
  \end{multline}
  for the dispersion function (\ref{D-function}), where $a$ is the
  lattice constant.  We adopt this form for the mobile particles in
  the Falicov-Kimball model;
  $D^\flavorf_{\alpha\beta}$ $=$ $0$ for the immobile species.
  
  \end{appendix}

  \vspace*{\fill}


\begin{thebibliography}{99}
\bibitem{footnote1}
    The current is gauge-invariant and satisfies the continuity
    equation for the density $\rho(\bmr) = \sum_{i\sigma}
    \delta(\bmr-\bmR_i)
    c_{i\sigma}^{\dagger}c_{i\sigma}^{\phantom{\dagger}}$, as under a
    gauge transformation the Hamiltonian transforms as $H\{\bmA+{\bm
      \nabla} \Lambda\}$ $=$ $e^{-ig} H\{\bmA\}e^{ig}$, where
    $g=\frac{e}{\hbar c}\int d^d\bmr \Lambda(\bmr)\rho(\bmr)$.
\end{thebibliography}
\end{document}